\documentclass[10pt,letterpaper]{article}
\usepackage{opex3}

\usepackage{graphicx}
\graphicspath{{pict/}{}}

\usepackage{amssymb}
\usepackage{bm}
\usepackage{here}

\newcommand{\be}{\begin{equation}}
\newcommand{\ee}{\end{equation}}

\newcommand\pictc[5]{\begin{figure}
                   \centerline{
                   \includegraphics[width=#1\columnwidth]{#3}}
               \protect\caption{\protect\label{fig:#4} #5}
                \end{figure}            }

\newcommand\pict[4][1]{\pictc{#1}{!tb}{#2}{#3}{#4}}
\newcommand\rpict[1]{\ref{fig:#1}}

\newcommand\leqt[1]{\protect\label{eq:#1}}

\newcommand\lsect[1]{\protect\label{sect:#1}}
\newcommand\rsect[1]{\ref{sect:#1}}

\newcounter{Fig}

\begin{document}
\begin{sloppy}
\title{Dispersion extraction with near-field measurements in periodic waveguides}

\author{Andrey A. Sukhorukov$^{\ast}$, Sangwoo Ha, Ilya V. Shadrivov, David~A.~Powell, and Yuri S. Kivshar}

\address{
Centre for Ultrahigh-bandwidth Devices for Optical Systems (CUDOS) and
Nonlinear Physics Centre, Research School of Physics and Engineering,
Australian National University, Canberra, ACT 0200, Australia\\
$^{\ast}$ Email: ans124@rsphysse.anu.edu.au
}

\begin{abstract}
We formulate and demonstrate experimentally the high-resolution spectral method based on Bloch-wave symmetry properties for extracting mode dispersion in periodic waveguides from measurements of near-field profiles. We characterize both the propagating and evanescent modes, and also determine the amplitudes of forward and backward waves in different waveguide configurations, with the estimated accuracy of several percent or less. Whereas the commonly employed spatial Fourier-transform (SFT) analysis provides the wavenumber resolution which is limited by the inverse length of the waveguide, we achieve precise dispersion extraction even for compact photonic structures.
\end{abstract}


\section{Introduction}

The latest advances in near-field measurements open new opportunities for experimental characterization of pulse dynamics inside complex photonic structures~\cite{Engelen:2007-401:NAPH}. 
The correlation measurements of the near-field probe signal and the reference pulse enable the determination of both amplitude and phase of the electric field at distinct time frames. These data enable direct visualization of pulse propagation in real space. Additionally, it was shown that spectra of the field profiles obtained through spatial Fourier-transform (SFT) can be used to reveal the dispersion of modes in optical waveguides~\cite{Gersen:2005-123901:PRL}.
Such analysis is especially valuable for photonic-crystal waveguides, where dispersion curves can possess complex features such as turning, inflection, and anti-crossing points~\cite{Gersen:2005-73903:PRL, LeThomas:2008-125301:PRB}. However, there exists {a fundamental limitation} on results obtained with SFT: $\Delta k \ge 2 \pi / L$, where $\Delta k$ is the resolution of the wavenumber, and $L$ is the structure length. This severely restricts the potential to determine the mode dispersion.

High-resolution spectral methods have been developed for analysis of temporal dynamics~\cite{Roy:1991-109:PRP, Mandelshtam:2001-159:RAR}, overcoming the limitations of Fourier-transform method. An implementation of such approach for dispersion extraction of photonic-crystal waveguide modes was recently shown with numerical simulations~\cite{Dastmalchi:2007-2915:OL}. In this work, we reveal that the spatial spectral analysis can be enhanced through the application of generic Bloch-waves symmetry properties in arbitrarily-shaped periodic dielectric waveguides. In particular, we show that the required structure size for dispersion extraction can be further reduced, offering additional advantages for characterization of short waveguide sections, which may form parts of compact photonic-crystal circuits.
Furthermore, we demonstrate accurate extraction of real and complex wavenumber values from near-field experimental measurements, precisely determining the mode dispersion both inside the transmission band and within the photonic band-gap regions.

The paper is organized as follows. In Sec.~\rsect{method} we formulate the general approach for single- and multi-mode periodic waveguides based on Bloch-wave symmetries. Then, in Sec.~\rsect{single} we demonstrate dispersion extraction for a single-mode periodic waveguide and transmission line, and in Sec.~\rsect{multi} present an example for two-mode homogeneous waveguide. The examples in Secs.~\rsect{single} and~\rsect{multi} are based on experimental measurements at the microwave frequencies, confirming the high accuracy and robustness of extraction procedure under the presence of noise in experimental data. Conclusions and outlook are presented in Sec.~\rsect{conclusions}.

\section{Dispersion extraction based on the Bloch-wave symmetries} \lsect{method}

The goal of high-resolution spectral analysis~\cite{Roy:1991-109:PRP} is to determine the frequencies or wavenumbers of a finite total number of modes ($M$) which primarily determine the system evolution. Let us consider how this methodology can be applied to extract dispersion for a periodic waveguide which supports $M$ modes in particular frequency range. Since each of the modes of a periodic waveguide satisfies the Bloch theorem~\cite{Joannopoulos:1995:PhotonicCrystals}, the complex electric field envelope of a waveguide mode with the index $m$ at the frequency $\omega$ can be expressed as $E(x,y,z; \omega) = \psi_m(x,y,z; \omega) \exp( i k_m z / d )$. Here $k_m$ is the Bloch wavenumber, $x$ and $y$ are the orthogonal directions transverse to the waveguide, $z$ is the direction of periodicity, $d$ is the waveguide period, and $\psi_m$ is the periodic Bloch-wave envelope function: $\psi_m(z) = \psi_m(z+d)$.
Then, the total field inside the waveguide can be presented as a linear superposition of propagating modes and radiative waves:
$E(x,y,z; \omega) = \sum_{m=1}^M a_m \psi_m(x,y,z; \omega)\exp( i k_m z / d ) + w(x,y,z;\omega)$, where $a_m$ are the mode amplitudes and $w(x,y,z;\omega)$ is the radiation field due to the excitation of non-guided waves.

The key question is how to use the general spectral methods~\cite{Roy:1991-109:PRP} to determine the mode amplitudes $a_m$ and wavenumbers $k_m$ which provide the best fitting for the experimental measurements of the electric field distribution. One approach is to use the periodicity property of Bloch-wave envelopes and represent them as infinite Fourier series, $\psi_m(x,y,z; \omega) = \sum_{s=-\infty}^{+\infty} \widetilde{\psi}_{m,s}(x,y;\omega) \exp( i 2 \pi s z / d)$. Then, by performing the measurements along a line with fixed transverse positions $(x_0,y_0)$, the problem is reduced to finding harmonics with the amplitudes $\widetilde{\psi}_{m,s}(x_0,y_0;\omega)$ and the corresponding wavenumbers $k_{m,s} = (k_m + 2 \pi s)$. In case of Bloch-waves with smooth profiles, terms with $s \ne 0$ are small allowing efficient dispersion extraction. It appears that the latter condition was satisfied in the analysis of Ref.~\cite{Dastmalchi:2007-2915:OL}, enabling accurate dispersion calculation. However, if the Bloch-wave profiles vary strongly over the unit cell, their spectrum would span a broad range of ${s}$, substantially complicating the spectral analysis.

We suggest that, in the general case of arbitrarily-shaped periodic waveguides with complex Bloch-wave profiles, accurate dispersion extraction can be performed using the electric field measurements at a set of $N$ periodic locations:
$U_n = E(x_0,y_0,z_0+n\;d; \omega)$, $n=1:N$.
The positions $(x_0,y_0)$ can be chosen at the locations with the maximum field amplitudes, reducing the effect of noise in measurements.
Taking into account the periodicity property of Bloch-wave envelopes $\psi_m$, we introduce new variables $A_m = a_m \psi_m(x_0,y_0,z_0;\omega) \exp( i k_m z_0 / d)$, and obtain a set of equations:
\begin{equation} \leqt{Un}
   U_n = \sum_{m=1}^M A_m \exp( i k_m n ) + w_n.
\end{equation}
Then, considering the radiation waves to be small, we can determine the Bloch-wave parameters which describe most accurately the measured field profile using established mathematical algorithms~\cite{Roy:1991-109:PRP, Marple:1987:DigitalSpectral}.
In general, the number of measurements shall be equal to or exceed the number of unknowns, $N \ge 2 M$.

We note that the spatial spectral analysis can benefit through the application of additional constraint due to the symmetry of forward and backward modes in dielectric waveguides, which wavenumbers are related as $k_{2m}=-k_{2m-1}$~\cite{Joannopoulos:1995:PhotonicCrystals}. Then, the required structure size for dispersion extraction is reduced: $N \ge 3 M / 2$. In order to take advantage of this relation, we use the least-squares method in place of general spectral algorithms~\cite{Roy:1991-109:PRP, Marple:1987:DigitalSpectral}. Specifically, we seek the values of $A_m$ and $k_m$ which minimize the functional
   $W = \sum_{n=1}^N |w_n|^2 / \sum_{n=1}^N |U_n|^2$.
For given wavenumbers, the minimum $W_A(\{k_m\}) = {\rm min}_A W$ is achieved when $\partial W / \partial A_m = \partial W / \partial A_m^\ast = 0$. It follows that the optimal amplitudes satisfy the linear matrix equation $C \cdot \widetilde{A} = B$, where components of vector $\widetilde{A}$ are the optimal amplitude values, components of the matrix $C$ are
$C_{pq} = \sum_{n} \exp[i (k_p - k_q^*) n]$, and vector $B$ components are
$B_p = \sum_{n} U_n \exp( -i k_p^* n)$ for $p,q = 1:M$. We can show that $W_A(\{k_m\}) = W_{A=\widetilde{A}} = 1 - \sum_p \widetilde{A}_p B_p^\ast / \sum_n |U_n|^2$. The remaining task is to find the absolute minimum $W_{\rm min} = \min_{k_m} W_A$ (note that, by definition, $W_A$ is real and positive), and this can be done numerically, for example by using the 'fminsearch' function in Matlab.

Due to the possible presence of radiation and noise in experimental measurements, the optimal value of the functional $W_{\rm min}$ will be larger than zero. Most importantly, this value can be used to estimate the accuracy of the dispersion extraction. Specifically, we define the confidence interval of wavenumbers as those corresponding to $W_A(k_m) < 2 W_{\rm min}$.

\section{Single-mode periodic waveguide} \lsect{single}

We now illustrate the application of the general approach for dispersion extraction in the regime when a periodic dielectric waveguide is single-moded. Then, we take $M=2$, accounting for one forward and one backward mode, which propagation constants are related as $k_2 = - k_1$. In this case, the required number of measurements at periodic locations for dispersion extraction is $N \ge 3$.
For three measurements ($N=3$) we can find an analytical solution to $W=0$: the wavenumbers are $k_1 = -k_2 = \cos^{-1}[ (U_1 + U_3) / (2 U_2)]$, and the corresponding amplitudes of counter-propagating waves are $\widetilde{A}_1 = [ 2 U_2 + i (U_1 - U_3) / \sin(k) ] / 4$ and $\widetilde{A}_2 = [ 2 U_2 + i (U_3 - U_1) / \sin(k) ] / 4$. The larger number of measurements allows one to estimate the accuracy due to the noise and radiation affecting experimental data. It is instructive to compare this procedure with the SFT, where the location of the spectral peaks can only be determined with accuracy no better than $\Delta k \ge 2 \pi / N$, and for $N=3$ practically no information on wavenumber values can be obtained.

\pict{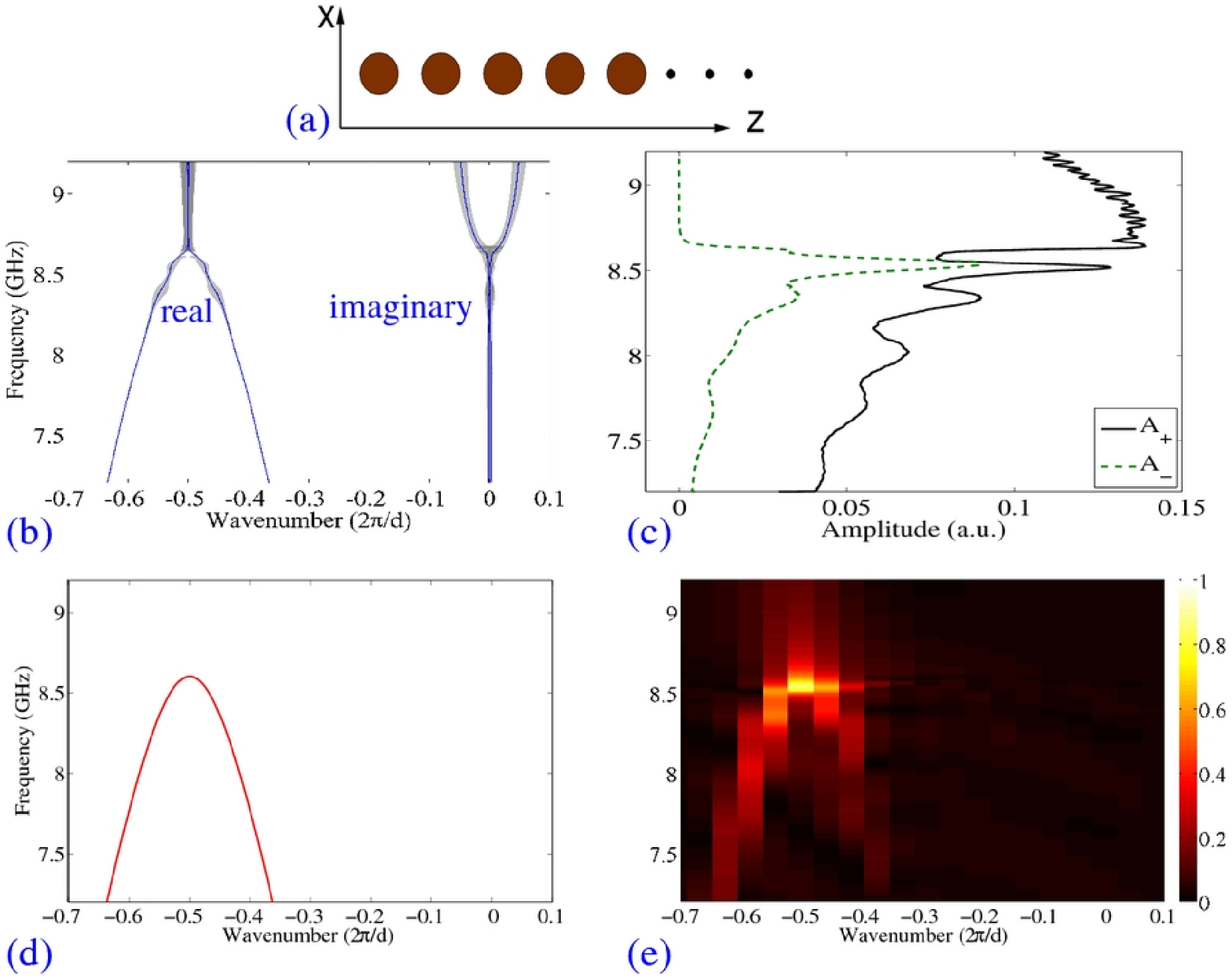}{rods}{
(a)~Sketch of the periodic waveguide made of dielectric pillars.
(b)~Dispersion retrieved from experimental measurements, shown are both the real and imaginary parts as indicated by labels. Grey shading marks the confidence interval.
(c)~Retrieved amplitudes of experimentally excited forward ($A_+$) and backward ($A_-$) waves.
(d)~Numerically calculated dispersion.
(e)~Normalized SFT of experimentally measured electric field profile at rod centers along the waveguide.
}

We now demonstrate dispersion extraction
from experimental data.
The measurements have been performed for microwave frequencies, where the amplitude and phase profiles of electric field distributions excited by vector network analyzer were sampled by a monopole near-field probe connected to the second port of the vector network analyzer.

First, we study wave propagation through an array of 20 dielectric pillars, as schematically shown in Fig.~\rpict{rods}(a). This system can be used as a testbed for the key features of wave propagation in periodic waveguides due to the presence of photonic band-gaps and associated changes in dispersion characteristics~\cite{Fan:1995-1267:JOSB}.
Each pillar is 1cm high, has a diameter of 4.18mm with the refractive index of 2.5, and they are sandwiched between the metal plates in the vertical direction. The field measurements are performed at each of the rod centers. The extracted dispersion for TM polarization is presented in Fig.~\rpict{rods}(b). We see that the essential features of the photonic band-gap are fully recovered. As the frequency is increased, the wavenumber approaches the edge of the Brillouin zone. Then, the imaginary part increases, indicating the exponential decay of the wave due to the photonic band-gap. Note that, close to the gap-edge, the real and imaginary parts of the wavenumber demonstrate the square-root dependence on frequency detuning, in agreement with the general properties of periodic photonic structures~\cite{Joannopoulos:1995:PhotonicCrystals}. Additionally, we also extract the amplitudes of Bloch waves which are excited experimentally, see Fig.~\rpict{rods}(c). The nontrivial dependence of amplitudes on the frequency is due to the
reflections at the structure boundaries. It is interesting to see that only one Bloch wave is present for a range of frequencies inside the band-gap, since the mode decays quickly and cannot excite the other evanescent wave at the opposite end of the waveguide. We also calculate numerically the modal dispersion considering the two-dimensional photonic structure with the effective refractive index of 2.4296, by computing eigenmodes of Maxwell's equations using the freely available MIT Photonic Bands (MPB) software package utilizing the plane-wave expansion method~\cite{Johnson:2001-173:OE}, see Fig.~\rpict{rods}(d). Our experimentally extracted dispersion curves perfectly agree with the calculated dependencies. For comparison, we also present the SFT spectrum of the electric field measurements in Fig.~\rpict{rods}(e). This spectrum can only indicate the general dispersion trend with very rough wavenumber resolution, and only inside the transmission band.

\pict{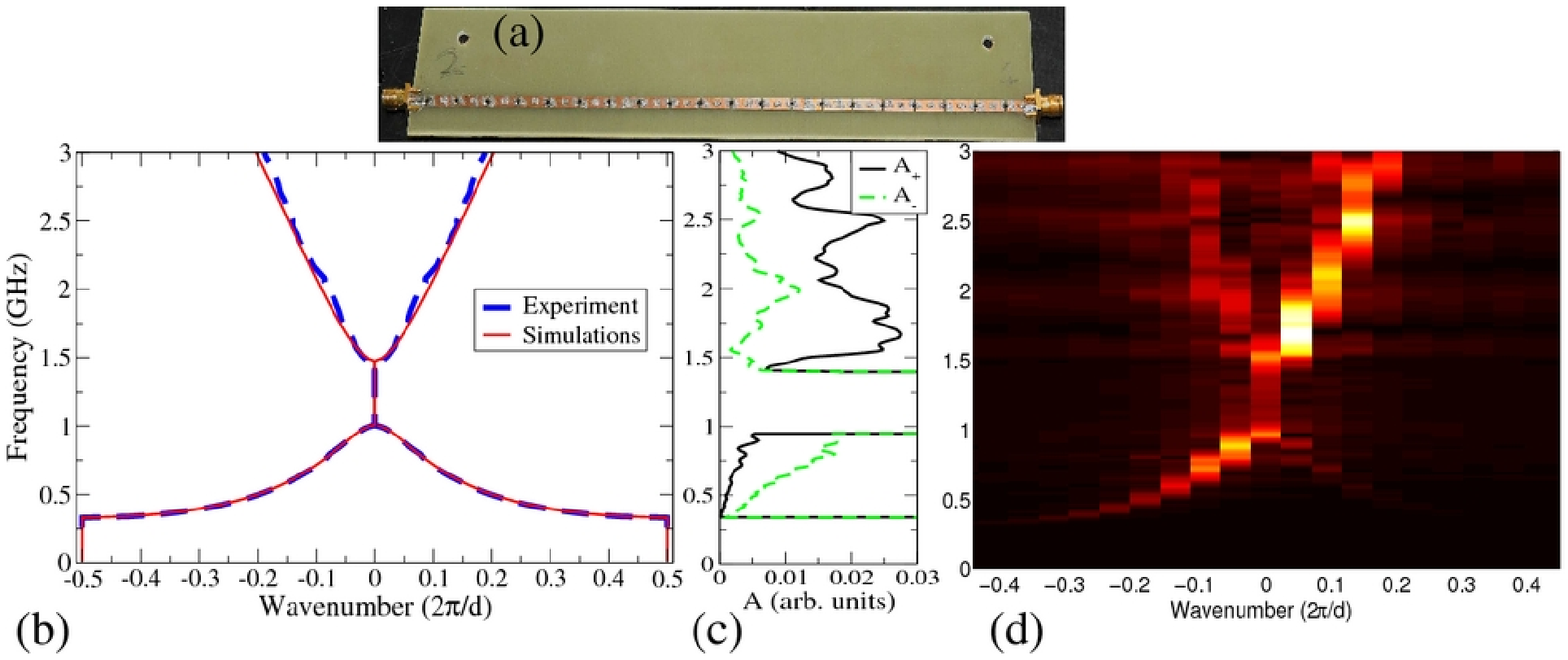}{transmLine}{
(a)~Photograph of transmission line.
(b)~Dispersion retrieved from experimental measurements (dashed lines) and calculated numerically (solid lines).
(c)~Retrieved amplitudes of experimentally excited modes with positive ($A_+$) and negative ($A_-$) wavenumbers.
(d)~SFT of experimentally measured electric field profile along the line.
}

Now we analyze the dispersion properties of the left-handed transmission line shown in Fig.~\rpict{transmLine}(a), which characteristics and nonlinear response were previously considered in Ref.~\cite{Powell:2008-264104:APL}. It was shown that for this structure, $\omega(k) = \omega(-k)$ similar to dielectric waveguides, and therefore we use this symmetry in dispersion extraction.
The line has 20 periods to ensure that end effects do not dominate the response, and the field distribution is registered at periodic locations along the line.
We observe an excellent performance of extraction method for this periodic structure, as demonstrated by close agreement of dispersion curves based on experimental data and calculated numerically, see Fig.~\rpict{transmLine}(b). We also determine the mode amplitudes and find that the dominant propagating mode has negative wavenumber at lower frequencies, and positive wavenumber at higher frequencies above the stop-band. This result agrees with the nature of modes in the transmission line~\cite{Powell:2008-264104:APL}. As in the previous example, we find that very limited information on mode dispersion can be obtained with SFT, shown in Fig~\rpict{transmLine}(c).

\section{Multi-mode waveguide} \lsect{multi}

We now demonstrate the dispersion extraction for a multi-mode waveguide. Specifically, we consider a homogeneous (non-periodic) perspex dielectric slab waveguide, shown schematically in Fig.~\rpict{twoMode}(a). The waveguide is 3cm wide and 20cm long, and it is sandwiched between the metal plates in the vertical direction. Under such conditions, the dispersion of TE modes can be calculated using expressions for planar waveguides~\cite{Born:2002:PrinciplesOptics} with the effective dielectric constant, which value for our system is found to be $\varepsilon_{\rm eff} = 2.1$.  In Fig.~\rpict{twoMode}(b) we show the numerically calculated dispersion for the first two modes with solid lines, presented for both forward- and backward-propagating waves with positive and negative wavenumbers, respectively.

In experiment, we measure the electric field amplitude at $N=71$ points with $d=2$mm spacing along the waveguide. The transverse antenna position is shifted away from the waveguide center by $x_0 = 1$cm, as indicated by dashed line in Fig.~\rpict{twoMode}(a). This position is chosen since at the waveguide center the amplitude of the second mode vanishes as its profile is anti-symmetric. Then, we process the data by running the retrieval algorithm for $M=4$ total number of modes, and we analyze the domain of real-valued wavenumbers $k_j$ in order to characterize the guided waves. The results of dispersion extraction are shown with dashed lines in Fig.~\rpict{twoMode}(b), revealing excellent agreement with numerical calculations. Indeed, we observe that for all frequencies the relative error $W_{\rm min}$ remains below 8\%.

For comparison, in Fig.~\rpict{twoMode}(c) we show the SFT of electric field amplitudes. The spectral resolution of this plot is severely limited due to the small waveguide length. Additionally, the amplitude of the second excited mode is much smaller than the fundamental mode, and even the existence of the second mode cannot be reliably determined from the Fourier spectra

\pict{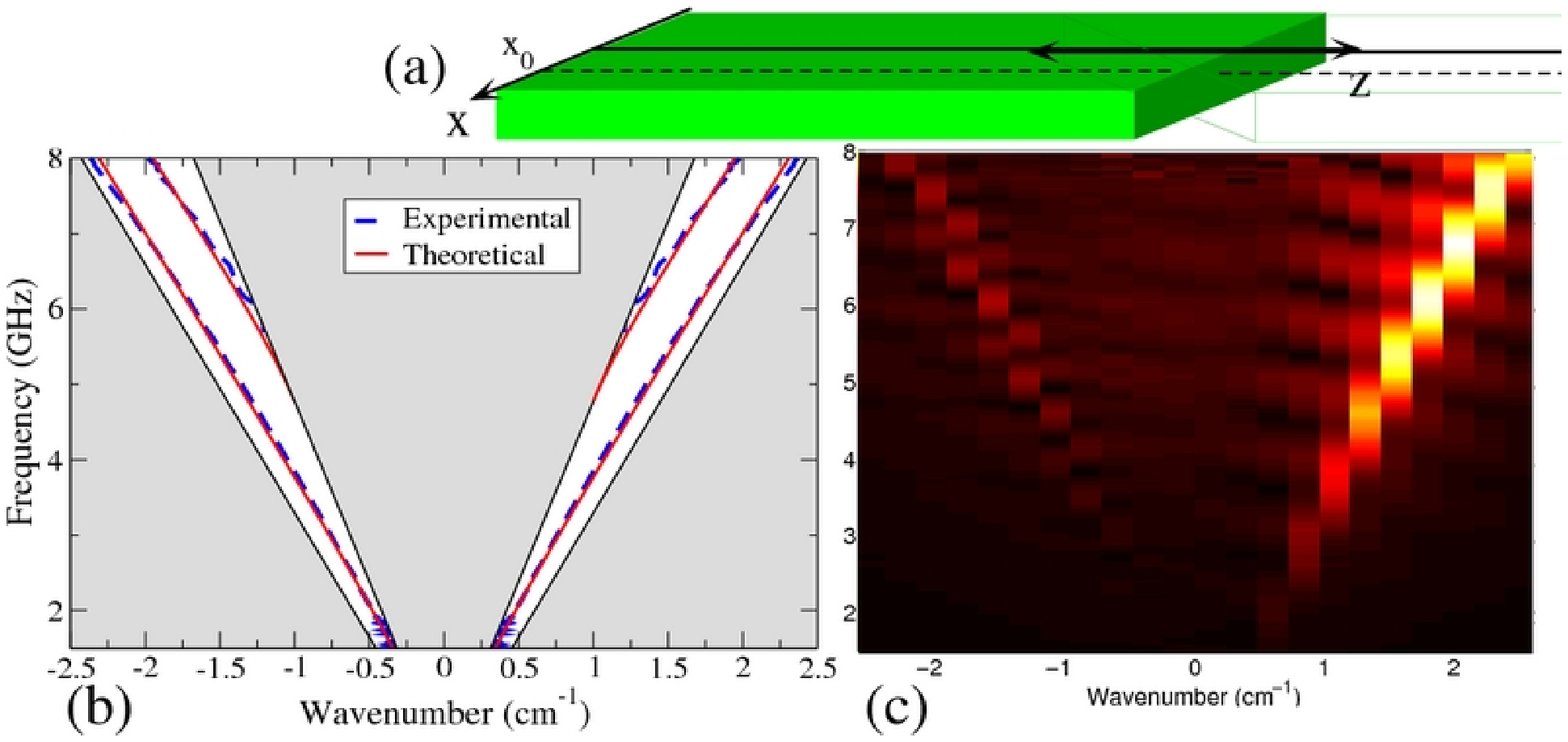}{twoMode}{
(a)~Sketch of perspex dielectric slab waveguide.
(b)~Dispersion retrieved from experimental measurements (dashed lines) and calculated numerically (solid lines). Shaded are the regions above the free-space light line, and below dielectric light line, where
guided modes do not exist.
(c)~SFT of experimentally measured electric field profile along the waveguide.
}

\section{Conclusion and outlook} \lsect{conclusions}

In conclusion, we have presented a generic approach based on high-resolution spectral methods combined with specific Bloch-wave symmetries for the extraction of mode dispersion and amplitudes using near-field measurements in periodic waveguides.
We have demonstrated its successful application through processing of experimental measurements for a variety of waveguide configurations, observing very good accuracy and high robustness to experimental noise.
One interesting problem for the future studies is to explore the possibility to extend this approach for two-dimensional Bloch waves.

\end{sloppy}
\end{document}